\documentstyle[aps,epsfig,a4]{revtex}
\begin{document}
\draft

\title{Evaluations of freeze-out parameters from ${ {dE_{T}} \over
{d\eta} } / { {dN_{ch}} \over {d\eta} }$ ratio measured at RHIC
and SPS }
\author{Dariusz Prorok}
\address{Institute of Theoretical Physics, University of
Wroc{\l}aw,\\ Pl.Maksa Borna 9, 50-204  Wroc{\l}aw, Poland}
\date{December 6, 2002}
\maketitle
\begin{abstract}
In the presented paper curves of constant $\epsilon_{T} /
n_{charged}$ are calculated in $T-\mu_{B}$ plane, in the framework
of a single-freeze-out thermal hadron gas model. The ratio is a
theoretical equivalent of ${ {dE_{T}} \over {d\eta} }_{\mid
\eta=0} / { {dN_{ch}} \over {d\eta} }_{\mid \eta=0}$ measured at
RHIC and SPS. In both $\epsilon_{T}$ and  $n_{charged}$ decays of
hadron resonances are taken into account. The freeze-out
temperature $T_{f.o.}=156_{-11}^{+14}$ MeV is obtained for RHIC,
whereas $T_{f.o.}=134-140$ MeV is evaluated for SPS.
\end{abstract}


In this letter, the allowed region of freeze-out parameters in
$T-\mu_{B}$ plane is established on the basis of ${ {dE_{T}} \over
{d\eta} }_{\mid \eta=0} / { {dN_{ch}} \over {d\eta} }_{\mid
\eta=0}$ ratio measured for Au-Au collisions at
$\sqrt{s_{NN}}=130$  GeV (RHIC) \cite{Adcox:2001ry} and Pb-Pb
collisions at $\sqrt{s_{NN}}=17.2$  GeV (SPS)
\cite{Aggarwal:2000bc}. Surprisingly, the ratio is roughly
constant as a function of the impact parameter (besides the narrow
region of peripheral collisions) and equals about $0.8$ GeV in
both experiments.

The concept of the presented paper is similar to the idea of
\cite{Cleymans:1998fq}, but here the information about freeze-out
conditions is extracted straightforward from the experimentally
measured quantity. In \cite{Cleymans:1998fq}, a curve of the
freeze-out in $T-\mu_{B}$ plane is obtained from the observation
that the average energy per hadron, calculated in the framework of
a thermal model, equals $1$ GeV at the chemical freeze-outs
determined for SPS, AGS and SIS. In the presented paper, the
thermal model with single freeze-out is used
\cite{Florkowski:2001fp,Broniowski:2001we,Broniowski:2001uk,Michalec:2001um}.
The model has been applied successfully to reproduce ratios and
$p_{T}$ spectra of particles observed at RHIC
\cite{Florkowski:2001fp,Broniowski:2001we,Broniowski:2001uk}. The
main assumption of the model is the simultaneous occurrence of
chemical and thermal freeze-outs. The new data on $K^{*}(892)^{0}$
production revealed by the STAR Collaboration \cite{Adler:2002sw}
support strongly this assumption. Since ${ {dE_{T}} \over {d\eta}
}_{\mid \eta=0} / { {dN_{ch}} \over {d\eta} }_{\mid \eta=0}$
measurement has been done at midrapidity, the presented
estimations of freeze-out parameters are valid for the Central
Rapidity Region (CRR) of heavy-ion collisions under consideration.

Therefore, it is assumed that a noninteracting gas of stable
hadrons and resonances at chemical and thermal equilibrium is
present at the CRR. As a first approximation, a static fireball is
considered (the incorporation of the expansion into the model will
be the subject of the subsequent paper). Then the distributions of
various species of primordial particles are given by usual
Bose-Einstein and Fermi-Dirac formulae. Only baryon number
$\mu_{B}$ and strangeness $\mu_{S}$ chemical potentials are taken
into account, here. The isospin chemical potential $\mu_{I_{3}}$
has very low value in collision cases considered
\cite{Florkowski:2001fp,Braun-Munzinger:1999qy}) and therefore can
be neglected. For given $T$ and $\mu_{B}$, $\mu_{S}$ is determined
from the requirement that the overall strangeness of the gas
equals zero. In this way, the temperature $T$ and the baryon
chemical potential $\mu_{B}$ are the only independent parameters
of the model.

At the temperature $T$ and the baryon chemical potential
$\mu_{B}$, the transverse energy density $\epsilon_{T}^{i}$ of
specie $i$ could be defined as ($\hbar=c=1$ always)

\begin{equation}
\epsilon_{T}^{i}= (2s_{i}+1)
\int_{-\infty}^{\infty}dp_{z}\;\int_{0}^{\infty}dp_{T}\;p_{T}\sqrt{m_{i}^{2}+p_{T}^{2}}\;
f_{i}(p;T,\mu_{B})\;, \label{energyt}
\end{equation}

\noindent where $p = \sqrt{p_{T}^{2} + p_{z}^{2}}$ and $m_{i}$,
$s_{i}$ and $f_{i}(p;T,\mu_{B})$ are the mass, spin and the
momentum distribution of the specie, respectively. At the
freeze-out the thermal system ceases and there are only freely
escaping particles instead of the fireball. The measured
${{dE_{T}} \over {d\eta} }_{\mid \eta=0}$ is fed from two sources:
(a) stable hadrons which survived until catching in a detector,
(b) secondaries produced by decays and sequential decays of
primordial resonances after the freeze-out. Therefore, if the
contribution to the transverse energy from particles (a) is
described, the distribution $f_{i}$ in (\ref{energyt}) is the
Bose-Einstein or Fermi-Dirac distribution at the freeze-out. But
if the contribution from particles (b) is considered, the
distribution $f_{i}$ is the spectrum of the finally detected
secondary and could be obtained from elementary kinematics of a
many-body decay or from the superposition of two or more such
decays (for details, see
\cite{Florkowski:2001fp,Broniowski:2001uk}; also
\cite{Sollfrank:1990qz,DeWit:it} could be very useful). In fact,
if one considers a detected specie $i$, then $f_{i}$ is the sum of
final spectra of $i$ resulting from a single decay (cascade) over
all such decays (cascades) of resonances that at least one of the
final secondaries is of the kind $i$.

Since measured ${ {dN_{ch}} \over {d\eta} }_{\mid \eta=0}$ has
also its origin in the above-mentioned sources (a) and (b), to
properly define the density of charged particles, decays should be
also taken into account. Thus, the density of measured charged
particle $j$ reads

\begin{equation}
n^{j} = n_{primordial}^{j} + \sum_{i} \alpha(j,i)\;
n_{primordial}^{i} \;, \label{nchj}
\end{equation}

\noindent where $n_{primordial}^{i}$ is the density of specie $i$
at the freeze-out and $\alpha(j,i)$ is the final number of specie
$j$ which can be received from all possible simple or sequential
decays of particle $i$. The density $n_{primordial}^{i}$ is given
by the usual integral of the Bose-Einstein or Fermi-Dirac
distribution.

Now, the theoretical equivalent of ${ {dE_{T}} \over {d\eta}
}_{\mid \eta=0} / { {dN_{ch}} \over {d\eta} }_{\mid \eta=0}$ could
be defined as

\begin{equation}
{{{ {dE_{T}} \over {d\eta} }_{\mid \eta=0}} \over {{ {dN_{ch}}
\over {d\eta} }_{\mid \eta=0}}} \equiv { {\epsilon_{T}} \over
{n_{charged}} } \;, \label{thratio}
\end{equation}

\noindent where the transverse energy density $\epsilon_{T}$ and
the density of charged particles $n_{charged}$ are given by

\begin{equation}
\epsilon_{T} = \sum_{i \in A} \epsilon_{T}^{i} \;, \label{enertot}
\end{equation}

\begin{equation}
n_{charged} = \sum_{j \in B} n^{j} \;. \label{nchtot}
\end{equation}

\noindent Note that there are two different sets of final
particles, $A$ and $B$ ($B \subset A$). $B$ denotes final charged
particles and these are $\pi^{+},\; \pi^{-},\; K^{+},\; K^{-},\;
p$ and $\bar{p}\;$, whereas $A$ also includes $\gamma,\;
K_{L}^{0},\; n$ and $\bar{n}\;$ \cite{Adcox:2001ry}. The formula
(\ref{enertot}) is the natural application (and generalization for
a thermal system) of the transverse energy definition from
\cite{Albrecht:1991fg}.

\begin{figure}
\begin{center}{
{\epsfig{file=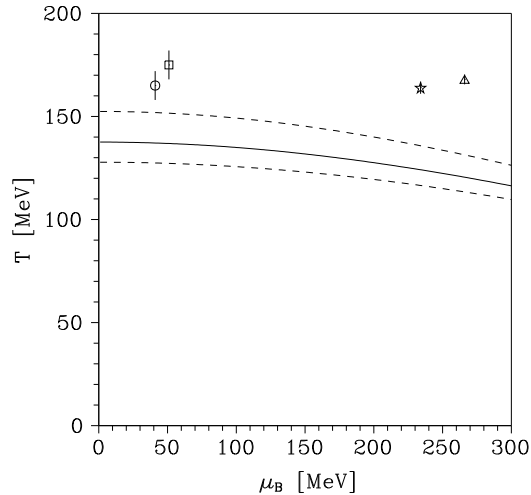,width=7cm}} }\end{center}
\caption{Curves of constant $\epsilon_{T} / n_{charged}$ for
$\epsilon_{T} / n_{charged}=0.8$ GeV (solid) and $\epsilon_{T} /
n_{charged}=0.74,\;0.88$ GeV  (dashed). Calculations have been
done for the excluded volume hadron gas model with a hard core
radius of $0.4$ fm and decays have not been included in
$\epsilon_{T}$. Also estimates of freeze-out parameters from
\protect\cite{Braun-Munzinger:2001ip} (square),
\protect\cite{Florkowski:2001fp} (circle),
\protect\cite{Michalec:2001um} (star) and
\protect\cite{Braun-Munzinger:1999qy} (triangle) are depicted.}
\label{Fig.1.}
\end{figure}

As the first step, the case of $\epsilon_{T}$ without decays
included is considered. This means that the transverse energy
density is given also by (\ref{enertot}) but with the sum over all
constituents of the gas and $f_{i}$ in (\ref{energyt}) is the
usual Bose-Einstein or Fermi-Dirac distribution at the freeze-out.
Following \cite{Braun-Munzinger:1999qy,Braun-Munzinger:2001ip},
the excluded volume hadron gas model is used. The foundations of
the model could be found in
\cite{Hagedorn:1978kc,Hagedorn:1980kb,Yen:1997rv} (in the
following, the formulation of \cite{Yen:1997rv} is used). The gas
consists of all mesons up to $K_{2}^{*}$ and baryons up to
$\Omega^{-}$ with antiparticles treated as a different specie
($151$ species all together). A hard core radius of $0.4$ fm is
put for all particles \cite{Braun-Munzinger:2001ip} (it has been
checked that for the radius equal to $0.3$ fm, the results are the
same). The results of calculations are presented in
Fig.\,\ref{Fig.1.}. The solid curve represents (\ref{thratio})
equal to $0.8$ GeV, whereas dashed curves are for the maximum and
minimum values of (\ref{thratio}) taken at the edges of error bars
of the last two points of the PHENIX measurement (see Fig.4b of
\cite{Adcox:2001ry}).

For comparison with previous estimates of the freeze-out
conditions, those results are also depicted as separate points.
They were obtained from thermal model fitting to the measured
particle ratios. The square denotes $T = 175 \pm 7$ MeV and
$\mu_{B} = 51 \pm 6$ MeV \cite{Braun-Munzinger:2001ip}, the circle
$T = 165 \pm 7$ MeV and $\mu_{B} = 41 \pm 5$ MeV
\cite{Florkowski:2001fp} and both are for RHIC. The star is at $T
= 164 \pm 3$ MeV and $\mu_{B} = 234 \pm 7$ MeV
\cite{Michalec:2001um}, the triangle at $T = 168 \pm 2.4$ MeV and
$\mu_{B} = 266 \pm 5$ MeV \cite{Braun-Munzinger:1999qy} and both
represent SPS conditions.

\begin{figure}
\begin{center}{
{\epsfig{file=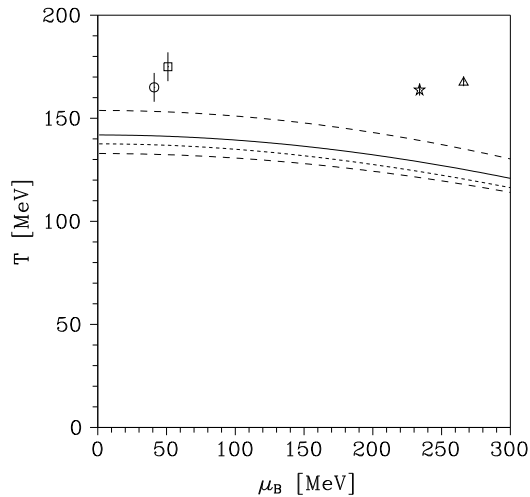,width=7cm}} }\end{center} \caption{Same
as Fig.\,\ref{Fig.1.} but for the gas consisting of 36 species
only. For comparison, the solid curve from Fig.\,\ref{Fig.1.} is
also repeated (short-dashed here).}
\label{Fig.2.}
\end{figure}

Since the aim of the presented paper is to calculate curves of
constant ratio (\ref{thratio}) with decays taken into account in
both $\epsilon_{T}$ and $n_{charged}$, some simplifications are
necessary. This is because the complete treatment of resonance
decays in $\epsilon_{T}$ is complex and consuming a lot of
computer working time in numerical calculations. Therefore the
initial set of resonances should be as small as possible. The
lifetime of at least $10$ fm is chosen as the necessary condition,
here. This reduces constituents of the hadron gas to $36$ species.
Of course, the condition is arbitrary but makes sense because: the
first, most neglected resonances have the lifetime of the order of
a few fm, so they could be thought of as decaying already at the
pre-equilibrium stage; the second, they have masses of the order
of $1$ GeV or more and because of the damping factor proportional
to $\exp{(-m/T)}$ their contribution to $\epsilon_{T}$ is
negligible in comparison with the lighter particles in the
temperature range considered here. The results of calculations of
curves of constant $\epsilon_{T} / n_{charged}$ ratio with no
decays included in the numerator and for the reduced gas case are
depicted in Fig.\,\ref{Fig.2.}. To compare with the previous case
of $151$ species, also the solid curve from Fig.\,\ref{Fig.1.} is
repeated as the short-dashed one. It can be seen that the
reduction in the number of species has not changed the results
substantially. Actually, the temperature has increased from
$T=137.6$ MeV to $T=141.9$ Mev (at $\mu_{B} = 1$ MeV), i.e. about
$3 \%$ for $\epsilon_{T} / n_{charged}=0.8$ GeV.

It has been also checked that the excluded volume corrections do
not change any of the presented results. The curves of constant
$\epsilon_{T} / n_{charged}$ are exactly the same for the excluded
volume hadron gas model and for the corresponding point-like one.
There are two reasons for that: the first, the volume corrections
placed in denominators of expressions for various densities cancel
each other in a ratio; the second, the eigenvolume of a hadron and
the pressure in the region of $T$ and $\mu_{B}$ considered are so
small that their product correction to the chemical potential is
negligible there. Therefore, the point-like non-interacting hadron
gas with $36$ species will be used in final calculations.

\begin{figure}
\begin{center}{
{\epsfig{file=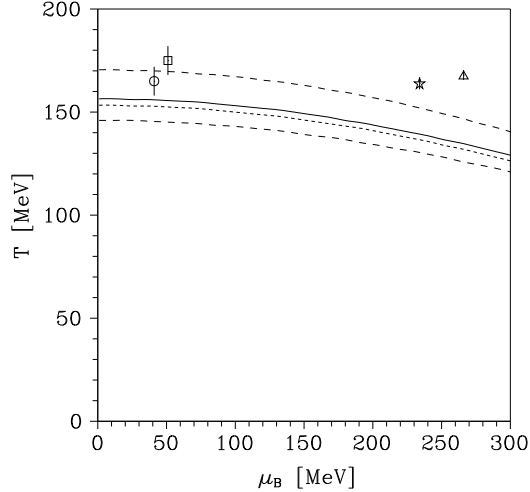,width=7cm}} }\end{center}
\caption{Curves of constant $\epsilon_{T} / n_{charged}$ for
$\epsilon_{T} / n_{charged}=0.8$ GeV (solid) and $\epsilon_{T} /
n_{charged}=0.74,\;0.88$ GeV  (dashed). Calculations have been
done for the point-like $36$-specie hadron gas model with decays
included in evaluations of $\epsilon_{T}$. The short-dashed curve
represents the error resulting from the simplifications done (see
the text for more details). Also estimates of freeze-out
parameters from \protect\cite{Braun-Munzinger:2001ip} (square),
\protect\cite{Florkowski:2001fp} (circle),
\protect\cite{Michalec:2001um} (star) and
\protect\cite{Braun-Munzinger:1999qy} (triangle) are depicted.}
\label{Fig.3.}
\end{figure}

As it has been already mentioned, the main difficulty in complete
treatment of decays in numerical evaluation of $\epsilon_{T}$ is
their complexity. Therefore some further simplifications should be
done. First of all, some decays and cascades are neglected: (i)
four-body decays, (ii) superpositions of two three-body decays,
(iii) superpositions of two three-body and one two-body decays,
(iv) superpositions of four two-body decays, (v) some decays of
heavy resonances with very small branching ratios. Their maximal
contribution to $\epsilon_{T}$ has been evaluated at $1.8 \%$.
Also some additional numerical simplifications have been done and
they cause that actual $\epsilon_{T}$ could differ $0.5 \%$ at
most from the calculated one. Thus, the real $\epsilon_{T}$ for
the $36$-specie hadron gas could be $2.3 \%$ higher maximally than
its evaluation. Finally, the results of numerical calculations of
curves of constant $\epsilon_{T} / n_{charged}$ with decays taken
into account are presented in Fig.\,\ref{Fig.3.}. The solid and
dashed curves have the same meaning as in
Figs.\,\ref{Fig.1.}-\ref{Fig.2.}. The short-dashed curve
represents the above-mentioned error, i.e. it has been obtained by
simple replacement of evaluated $\epsilon_{T}$ with the value
$1.023 \cdot \epsilon_{T}$. This causes the {\it decrease} of the
temperature from $156.4$ MeV to $153.3$ MeV (at $\mu_{B} = 1$
MeV), i.e. about $2 \%$. But remember that the reduction in the
number of species has caused the {\it increase} of the temperature
by about $3 \%$. So, those two effects should cancel each other
and the solid and dashed curves of Fig.\,\ref{Fig.3.} reflect the
realistic physical conditions of the freeze-out in the RHIC and
SPS range.

From the solid curve of Fig.\,\ref{Fig.3.} the freeze-out
temperature could be found at $T_{f.o.} \simeq 156$ MeV for RHIC
and $T_{f.o.} \simeq 134-140$ MeV for SPS, if one puts the
corresponding baryon chemical potential values at estimates of
\cite{Florkowski:2001fp,Michalec:2001um,Braun-Munzinger:1999qy,Braun-Munzinger:2001ip}.
Similarly, the allowed range of the freeze-out temperature could
be established from dashed curves of Fig.\,\ref{Fig.3.} and one
has $T_{f.o.}=145-170$ MeV for RHIC and $T_{f.o.}=125-153$ MeV for
SPS. Note that for RHIC the good agreement with the prediction of
\cite{Florkowski:2001fp} has been obtained, also the estimate from
\cite{Braun-Munzinger:2001ip} agrees qualitatively. Unfortunately,
much worse agreement with the previous predictions for SPS
\cite{Michalec:2001um,Braun-Munzinger:1999qy} has been reached.
Nevertheless, these points could be just on the edge of the error
bar of SPS measurement, because SPS data have an additional $\pm
20 \%$ overall systematic error which is not shown in Fig.4b of
\cite{Adcox:2001ry} (and is not shown in Fig.\,\ref{Fig.3.},
neither). If one takes this additional error into account, then
the maximal possible value of ${ {dE_{T}} \over {d\eta} }_{\mid
\eta=0} / { {dN_{ch}} \over {d\eta} }_{\mid \eta=0}$ is about $1$
GeV for SPS \cite{Aggarwal:2000bc}. The estimate of
\cite{Michalec:2001um} gives $\epsilon_{T} / n_{charged}= 0.95$
GeV (within errors the minimal possible value is $0.93$ GeV), for
the case of \cite{Braun-Munzinger:1999qy} $\epsilon_{T} /
n_{charged}= 1$ GeV (within errors the minimal possible value is
$0.99$ GeV). Thus, the SPS data for ${ {dE_{T}} \over {d\eta}
}_{\mid \eta=0} / { {dN_{ch}} \over {d\eta} }_{\mid \eta=0}$ do
not contradict the thermal model predictions done in
\cite{Michalec:2001um,Braun-Munzinger:1999qy}, at least.

In conclusion, the region of the freeze-out parameters for RHIC
and SPS heavy-ion collisions has been established on the basis of
${ {dE_{T}} \over {d\eta} }_{\mid \eta=0} / { {dN_{ch}} \over
{d\eta} }_{\mid \eta=0}$ measurement
\cite{Adcox:2001ry,Aggarwal:2000bc}. The point-like
non-interacting hadron gas model with $36$ species has been used
in final calculations. Decays and sequential decays of
constituents of the gas have been taken into account there. The
good agreement with the previous estimates of freeze-out
conditions at RHIC \cite{Florkowski:2001fp,Braun-Munzinger:2001ip}
obtained from the analysis of measured particle ratios has been
found. Also predictions for SPS
\cite{Michalec:2001um,Braun-Munzinger:1999qy} can be accepted, but
with much worse accuracy. However, it should be stressed that
actually not the transverse masses are measured but energies times
$\sin{\theta}$ ($\theta$ is the polar angel). Also RHIC is the
opposite beam experiment whereas SPS is the fixed target one,
which should be taken into account in theoretical modelling. Both
effects could lower the estimated ratio substantially. The
considerations of the more realistic case will be the subject of
the next paper.

The author acknowledges very stimulating discussions with
Professor Ludwik Turko. This work was supported in part by the
Polish Committee for Scientific Research under Contract No. KBN -
2 P03B 030 18.

\end{document}